\documentclass[%
preprint,%
 onecolumn,
 amssymb, amsmath,%
 aip,apm,%
 superscriptaddress
]{revtex4-1}
\usepackage{graphicx}
\usepackage{dcolumn}%
\usepackage{array}
\usepackage{multirow}
\usepackage{overpic}
\usepackage{subfig}
\usepackage[table]{xcolor}
\usepackage{bm}%
\usepackage[colorlinks=true,linkcolor=blue]{hyperref}%
\expandafter\ifx\csname package@font\endcsname\relax\else
 \expandafter\expandafter
 \expandafter\usepackage
 \expandafter\expandafter
 \expandafter{\csname package@font\endcsname}%
\fi
\newcommand{\super}[1]{\ensuremath{^{\textrm{#1}}}}
\newcommand{\sub}[1]{\ensuremath{_{\textrm{#1}}}}
\newcommand{\tilda}[1]{\ensuremath{\sim{}}}

\colorlet{lightgray}{gray!18!white}
 \newcommand{\graycell}{\cellcolor{lightgray}}     
\begin{document}
\title[]{Localized electromechanical interactions in ferroelectric P(VDF-TrFE) nanowires investigated by scanning probe microscopy}
\author{Yonatan Calahorra}
\author{Richard A Whiter}
\author{Qingshen Jing}\affiliation{Department of Materials Science and Metallurgy, University of Cambridge, Cambridge CB3 0FS, United Kingdom}
\author{Vijay Narayan}\affiliation{Department of Physics, Cavendish Laboratory, J.J. Thompson Avenue, University of Cambridge, Cambridge CB3 0HE, United Kingdom}
\author{Sohini Kar-Narayan}\email{sk568@cam.ac.uk}
\affiliation{Department of Materials Science and Metallurgy, University of Cambridge, Cambridge CB3 0FS, United Kingdom}

\begin{abstract}
We investigate the electromechanical interactions in individual P(VDF-TrFE) nanowires in response to localized electrical poling via a conducting atomic force microscope tip. Spatially resolved measurements of piezoelectric coefficients and elastic moduli before and after poling reveal a striking dependence on the polarity of the poling field, notably absent in thin films of the same composition. These observations are attributed to the unclamped nature of the nanowires and the inherent asymmetry in their chemical and electrical interactions with the tip and underlying substrate. Our findings provide insights into the mechanism of poling/switching in polymer nanowires critical to ferroelectric device performance.\\
\end{abstract}%



\maketitle
\begin{figure}
\centering
\includegraphics[scale=1]{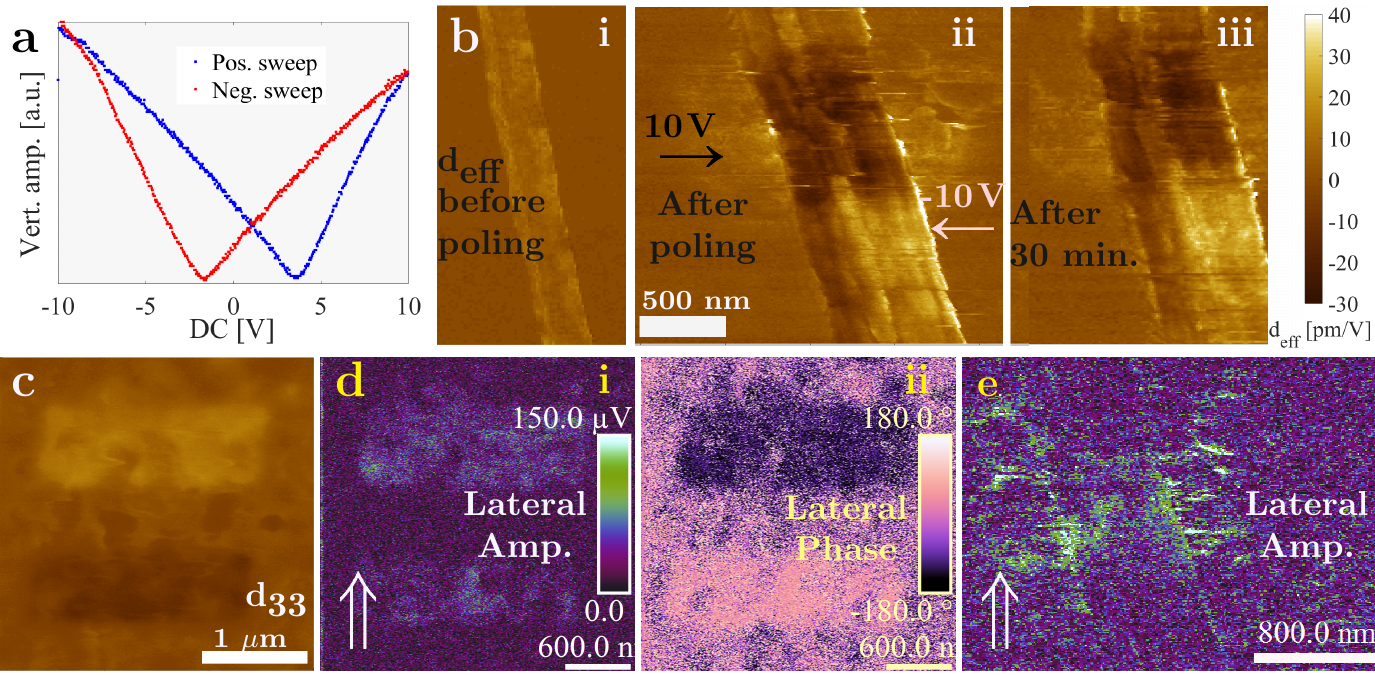}
\caption{a) A PE amplitude hysteresis loop obtained atop a NW; b) vertical PFM signals in pm/V from a NW subjected to local poling of $\pm$10 V: (i) before, (ii) immediately after poling, and (iii) 30 minutes after poling; c) vertical PFM signals in pm/V (same scale as in (b)), after local poling of the TF sample; d) lateral PFM (i) amplitude and (ii) phase obtained simultaneously with (c); e) lateral PFM amplitude from the NW, obtained simultaneously with (b)ii. The double arrow indicates the tip orientation during the scan.} 
\label{fig:poling}
\end{figure}
\begin{figure}
\centering
\includegraphics[scale=1]{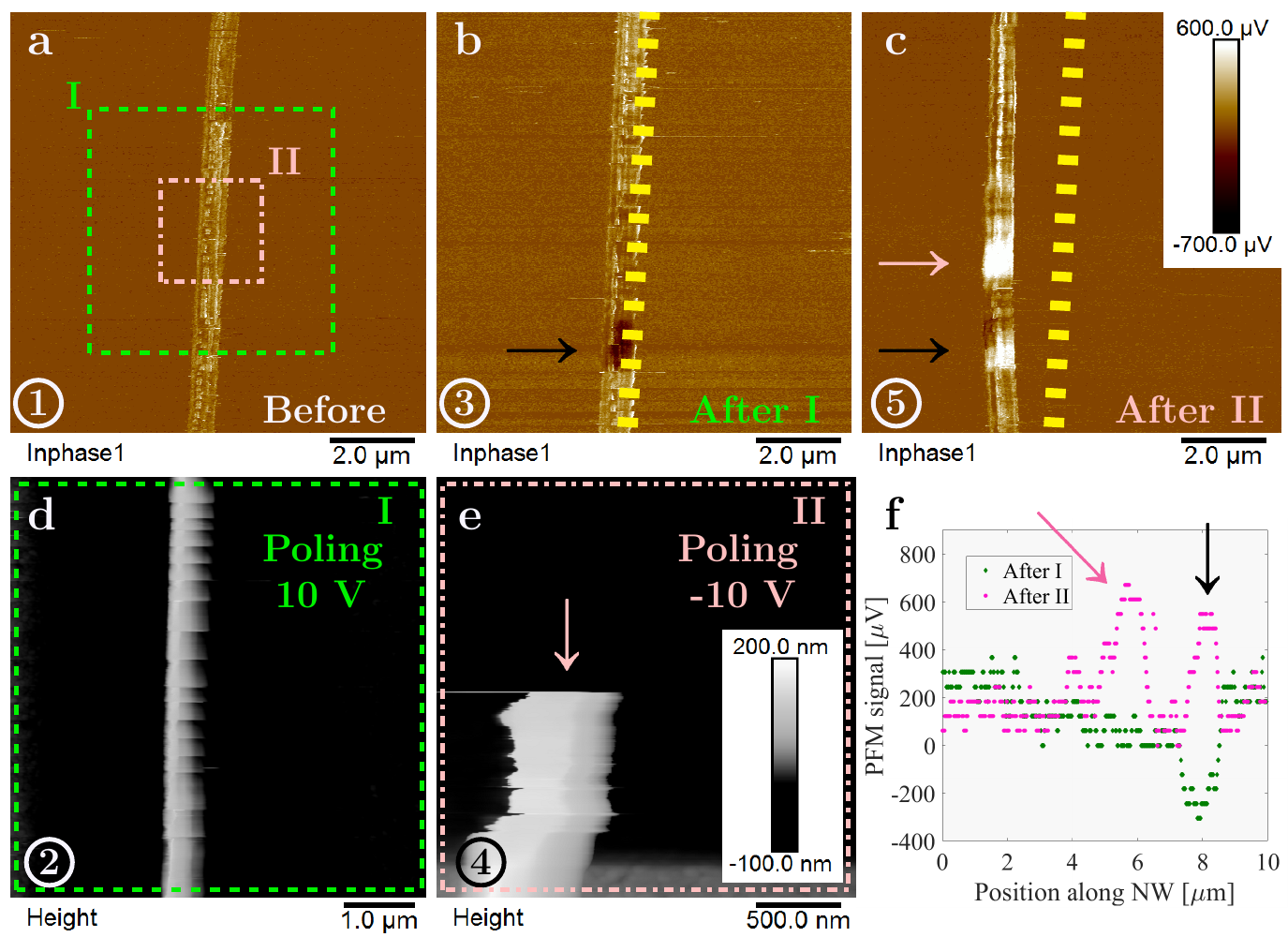}
\caption{Displacement of a single NW following a localized poling process. The numbering in the bottom left of the images indicates the capture sequence. Vertical PFM images (raw data) (a) before poling, (b) after up-poling, and (c) after down-poling; green and pink dashed squares depict up and down-poled areas respectively. Dashed yellow lines show original NW location; Topography of the NW during: (d) up- and (e) down-poling, whereby NW vanishes from the scanning frame. f) Vertical PFM cross-sections obtained from the NW in (b) and (c) (green and pink). Arrows indicate poled regions on the NW: black arrows for up-poled regions, notably switching contrast; and pink arrow for down-poled regions. }
\label{fig:movement}
\end{figure}
\begin{figure}
\centering
\includegraphics[scale=1]{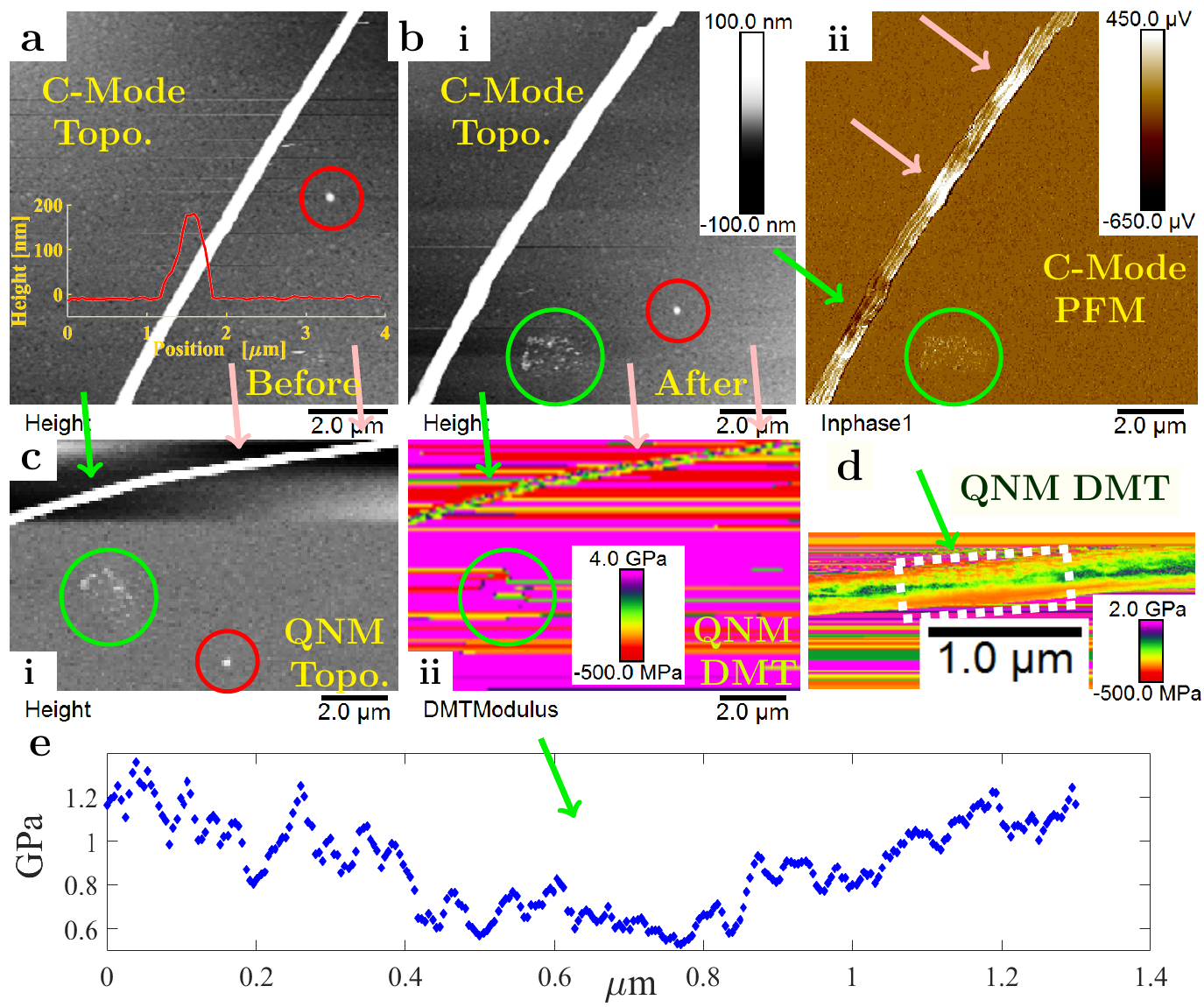}
\caption{Effect of poling on localized mechanical properties of a single NW: a) contact mode topography image of a NW before poling; b) (i) topography and (ii) PFM of the same NW following up- (green arrow) and down-poling (pink arrows); c) corresponding QNM mode (i) topography and (ii) modulus images from the NW; d) elastic modulus channel scanning the up-poled region; e) Line scan obtained from poled area (dashed box) in (d), showing the diminished modulus of the up-poled region. The red and green circles indicate the same features captured in different scans/channels, verifying the spatial location of the scans.}
\label{fig:QNM}
\end{figure}
\begin{figure}
\centering
\includegraphics[scale=1]{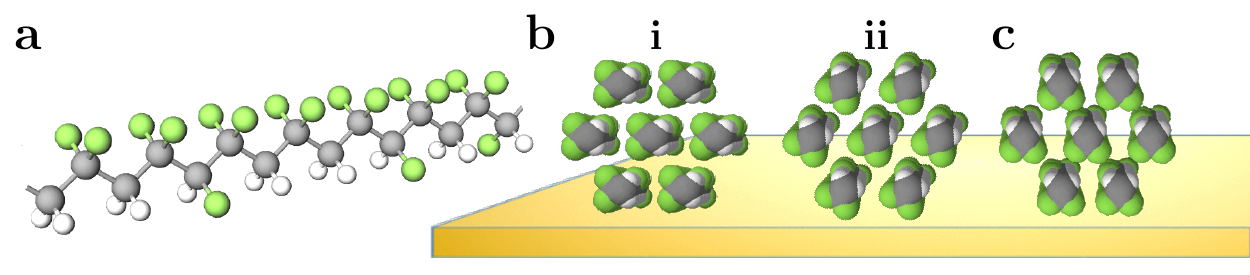}
\caption{Schematic illustrations of P(VDF-TrFE) chains: a) 3D view of a single all-trans chain, showing the different termination of the fluorine (larger green spheres) and hydrogen (smaller white spheres) edges of the chain; b) chain side view, with the (i) lateral and (ii) 60$^{\circ}$ rotated configurations, corresponding to (200) and (110) planes aligned with the surface; c) the theoretically unstable configuration of complete up-poled dipoles. (Image prepared with the help of \textit{http://molview.org}.)}
\label{fig:scheme}
\end{figure}

Polyvinylidene fluoride (PVDF) and its copolymers have been widely considered to be the materials of choice amongst ferroelectric (FE) and/or piezoelectric (PE) polymers, with applications in mechanical energy harvesters\cite{Whiter2014,Kim2013,Li2014apl,Kim2015}, memory devices\cite{Jonas2009}, sensors\cite{Persano2013,Sung2015}, and organic transistors\cite{Tian2016} as well as in medical applications due to their bio-compatibility\cite{Jung2016}. The FE/PE properties of these polymers arise due to the molecular dipoles present in the monomer unit (CH\sub{2}-CF\sub{2}) that are aligned perpendicular to the main chain axis\cite{Furukawa1989,Furukawa2006}. These semi-crystalline polymers are relatively cheap and easy to process, and are well-suited for applications requiring device flexibility. These properties often compensate for their lower piezoelectric coefficients as compared to more traditional ceramic FE/PE materials such as PZT (Pb[Zr\sub{x}Ti\sub{1-x}]O\sub{3}) and BTO (BaTiO\sub{3})\cite{Crossley2014,Crossley2015}. In comparison with the pure PVDF polymer, the co-polymer polyvinylidene fluoride-trifluoroethylene) (P(VDF-TrFE)) exhibits a higher occurrence of the FE $\beta$-phase, resulting in improved PE performance\cite{Li2014,Bohlen2014,Whiter2016}. However, the presence of the $\beta$-phase alone is insufficient to induce the PE character in these polymers as ``poling" is also required in order to orient the dipoles within the individual lamellae in the same direction, imposed via an externally applied electric field, or by mechanical stretching\cite{Furukawa1989,Cauda2015review}.\\ \indent
In recent years, it has been shown that P(VDF-TrFE) nanowires (NWs) grown via template-assisted fabrication methods\cite{Whiter2014} possess good PE properties without the need of high-voltage electrical poling and/or mechanical stretching procedures\cite{Kim2013nws,Jonas2013,Cauda2013nanoconfinement,Cauda2015review}. This is unlike nanowires grown by electrospinning where such procedures are inherent in the growth process\cite{Sencadas2012,Baniasadi2015,Baniasadi2016}. This ``self-poled" nature of template-grown P(VDF-TrFE) NWs along their axis makes them particularly attractive for PE nanogenerators for energy harvesting. We have recently demonstrated a working energy harvesting device based on as-grown P(VDF-TrFE) NWs\cite{Whiter2014}, fabricated within nanoporous anodized aluminum oxide (AAO) templates, and subsequently investigated the material properties of the NWs, to establish that the template-induced confinement during the growth process results in self-poled P(VDF-TrFE) NWs with enhanced PE properties\cite{Whiter2016}.\\ \indent
Studying materials at the nanoscale allows examination of fundamental properties and may subsequently contribute to achieving better device performance and control. Herein we present a scanning probe microscopy based study of individual P(VDF-TrFE) NWs that have been freed from the template in which they were grown. Notably, we employ for the first time, on the same location, both piezo-response force microscopy (PFM)\cite{Christman1998} to probe and control the PE and FE properties of single NWs dispersed on conducting substrates, as well as quantitative nanomechanical mapping (QNM, see Supporting Information Sec.\nolinebreak\ I for basics of operation)\cite{QNM} to study the resulting mechanical properties of these NWs, with respect to localized electrical manipulations. We find evidence that localized electrical poling of micrometre-sized NW sections, as achieved by applying an electric field across the thickness of the NW via a conductive atomic force microscope (AFM) tip, induced a structural and mechanical change to the material on a macromolecular level, in addition to the expected changes at the molecular level. While mechanical stress and/or strain are known to affect the FE properties of P(VDF-TrFE)\cite{Jonas2009,Song2015,Choi2015}, the opposite is rarely reported. Recently Baniasadi et al.\nolinebreak\ have observed a correlation between annealing procedures and electromechanical properties of electrospun P(VDF-TrFE) NWs\cite{Baniasadi2016}. Interestingly, we find that in our template-grown NWs, the local electromechanical interactions are polarity dependent, yielding different responses for poling ``up" and ``down"; i.e.\nolinebreak\ when the AFM tip is biased negatively (up) or positively (down) with respect to the substrate on which the NWs are dispersed (Supporting Information Figure S1). A nanowire within an SPM apparatus constitutes a highly localized and asymmetric system, thus allowing access to observations which are elusive in the traditional plate-capacitor configuration that is widely used to study polarization in FE materials\cite{Furukawa2006,Guo2013,Zeng2016}.\\ \indent
The full synthesis procedure of the NWs is identical to what has been previously reported\cite{Whiter2014}, and can also be found here in Supporting Information Sec.\nolinebreak\ II. Briefly, P(VDF-TrFE) powder (70:30) was dissolved in butan-2-ol to make a 10 wt\% solution which was dropped onto the AAO template with 200 nm nominal pore-diameter. Following infiltration and annealing at 60\super{o}C to form the NWs, the template was dissolved in phosphoric acid. The NWs (of diameter 140-200 nm as measured by AFM - see Fig.\nolinebreak \ref{fig:QNM}a) were then dispersed on a gold-coated silicon substrate by drop-casting from solution. Similarly, samples of thin P(VDF-TrFE) films of thickness $\sim$100 nm were prepared by spin-coating a 2 wt\% solution of identical polymer and solvent components on gold-coated silicon substrates and annealing at 100\super{o}C. Films were also prepared under similar annealing conditions as the NWs, i.e., annealed at 60\super{o}C for 24 hours, for a more thorough comparison. In all cases, the substrates were prepared by depositing 10~nm of Ti followed by 100~nm of Au by thermal evaporation at a pressure of less than 5$\times 10^{-6}$~mbar. Note that films having thickness similar to the NW diameter ($\sim$200~nm) were prepared as well, but the maximum possible voltage bias in our AFM setup was found to be insufficient for poling these films, as shown in Supporting Information Sec. III., and thus only the 100 nm-thick films were used for the localised poling studies. The samples were mounted on a 1 cm magnetic disc suitable for conductive AFM measurements using a \textit{Bruker} Multimode 8 (with Nanoscope V controller) microscope. Calibration for both PFM and QNM modes was performed using dedicated calibration samples (PFM-SMPL and PSFILM-12M by \textit{Bruker}) as described in Supporting Information Sec.\nolinebreak\ III. \\  \indent
Figure \ref{fig:poling}a shows the vertical PFM amplitude signal from the point where the AFM tip contacts an individual P(VDF-TrFE) NW during a DC voltage sweep (blue curve for up-poling, red curve for down-poling) with hysteresis indicative of FE switching/poling. Note that although the curves are not symmetrical around zero bias, the amplitude values for the different polarizations are relatively similar. This measurement was performed in continuous DC mode, and therefore a typical V-shape curve is observed due to increased electrostatic cantilever-sample interaction\cite{Hong2001}. Figures \ref{fig:poling}b \& c show the vertical PFM signal corresponding to the out-of-plane piezo-response of a NW before and after the application of $\pm$10 V across 1 $\mu$m sections of the NW (b) and thin film (TF) (c). The color and sign contrast are a clear indication of ferroelectricity in these NWs, and measurements 30 or 45 minutes after the poling process maintain the trend; e.g., Fig.\nolinebreak\ \ref{fig:poling}b (ii) and (iii) shows the same location imaged 30 minutes apart after poling treatment. Figure \ref{fig:poling}d shows {the uncalibrated lateral PFM signal amplitude (i) and phase (ii) recorded simultaneously with the corresponding vertical PFM signal for TF.} Figure \ref{fig:poling}e { shows the lateral PFM amplitude from the NW, obtained simultaneously with [b ii]}. The extracted results for the effective piezoelectric coefficient, \textit{d}\sub{eff}, from the NW and TF samples, with different tip-velocities during poling, and different times after the poling are presented in Table \ref{tbl:poling}.\\ \indent
\begin{table*}[ht]
\small
\center
  \caption{Extracted $|$\textit{d}\sub{eff}$|$ for P(VDF-TrFE) NWs and TFs. Gray tint indicates a measurement on the same location 30 minutes later. The voltage was applied to the sample, hence the up- and down-poled orientations.}
  \label{tbl:poling}
  \setlength{\extrarowheight}{2pt}
\begin{ruledtabular}\begin{tabular}{>{\centering\arraybackslash}m{0.12\textwidth}>{\centering\arraybackslash}m{0.16\textwidth}>{\centering\arraybackslash}m{0.16\textwidth}>{\centering\arraybackslash}m{0.16\textwidth}>{\centering\arraybackslash}m{0.16\textwidth}>{\centering\arraybackslash}m{0.16\textwidth}}
    Sample & Tip velocity [$\mu$m/s] & Un-poled [pm/V]& {\large $\downarrow$} (@ -10 V) [pm/V]& {\large $\uparrow$} (@ 10 V) [pm/V]& See \\ \hline \hline \vspace{5pt} 
 \multirow{2}{*}{TF (100\super{o}C)} & \multirow{2}{*}{2} & \tilda{}0  & 10.1$\pm$2.5 & 9.2$\pm$2.3 & Fig.\nolinebreak\ \ref{fig:poling}c\\ 
      &  & \graycell \tilda{}0  & \graycell 8.8$\pm$2.5 & \graycell 6$\pm$1.3 & \graycell \\ \hline
TF (60\super{o}C)& 2 & \tilda{}0 & 3.8$\pm$1.1 & 5.6$\pm$1.2 & Fig.\nolinebreak\ S4d\\ \hline
\multirow{2}{*}{NW} & \multirow{2}{*}{\ 2-3\footnotemark[1]} & 5.5$\pm$2.5 ($\downarrow$) & 19.6$\pm$4.6 & 13.4$\pm$5.8 & Fig.\nolinebreak\ \ref{fig:poling}b \\ 
      &  & \graycell 7.7$\pm$2.2 ($\downarrow$) & \graycell 21.6$\pm$4.1 & \graycell 15.5$\pm$6.2 & \graycell Fig.\nolinebreak\ \ref{fig:poling}b \\ \hline
\multirow{2}{*}{NW\footnotemark[2]} & \multirow{2}{*}{1.3} & 6$\pm$3.7 ($\downarrow$) & 32.1$\pm$3.3 & 9.8$\pm$4.2 & \\ 
      &  & \graycell 5.5$\pm$2.9 ($\downarrow$) & \graycell 27.1$\pm$2.3 & \graycell 9.1$\pm$3.5 & \graycell Fig.\nolinebreak\ S3a \\ 
  \end{tabular} \end{ruledtabular} \\ \flushleft
  \footnotetext[1]{This is an estimated velocity since data was not captured during poling}
  \footnotetext[2]{The measurements were taken 15 and 45 minutes after poling}
\end{table*}
Several important observations can be made following these experiments and from the results presented in Table \ref{tbl:poling}:(i) the pristine ``unpoled" regions of the TF and NW had different \textit{d}\sub{eff} values. While the films exhibited values close to zero as expected, the NWs exhibited a non-zero, “down”-poled oriented value. This can be attributed to the inherent self-poled nature of the NW, as has been discussed at length previously\cite{Whiter2016}. (ii) The efficiency of the poling procedure in NWs was found to be velocity-dependent and asymmetric with regards to “up”- and “down”- poling (with “down” poling usually stronger). While symmetric poling was achieved in the TF sample, NW poling remained asymmetric even with a slower scan. (iii) The poled regions in the TF gave rise to distinct lateral PFM signals that were mostly absent in poled NW regions (appearing only at domain boundaries). {The lateral signals exhibited opposite phases in the oppositely poled regions.} (iv) The \textit{d}\sub{eff} {values obtained from TFs are considerably smaller than from NWs.} (v) Some local NW damage resulting in PE-active residue being present on the passive substrate was only observed during up-poling of the NWs, as evident by the bright surface contrast around the dark area in Fig.\nolinebreak\ \ref{fig:poling}b (another example is the circled region in Fig.\nolinebreak\ \ref{fig:QNM} below). Generally, the \textit{d}\sub{eff} values reported here are comparable to other reports in the literature\cite{Choi2010nanoscale,Choi2015}, indicating adequate processing. The larger coefficients found for NWs are also in agreement with previous reports\cite{Jonas2013}.\\ \indent
Figure \ref{fig:movement} shows a series of AFM images depicting the $\mu$m-scale movement of a NW following tip-induced electrical poling. This localized movement of the NW relative to the substrate was also observed optically as shown in Supporting Information Figure S7. The section of the NW which moves is only a small fraction of the total length and is mostly restricted to the region where the poling process is undertaken.  Thermal drift and/or other imaging artifacts were ruled out as such movement was absent when the NWs were imaged without a poling field being applied. Figures \ref{fig:movement}a,b,c correspondingly show the vertical PFM signal of the NW before any poling process was undertaken, after an up-poling procedure, and after a subsequent down poling procedure, designed in a “sandwich” configuration (see designated poling areas in Fig.\nolinebreak\ \ref{fig:movement}a). Remarkably, the NW was found to exhibit a \tilda{}2 $\mu$m movement, brought about by the down-poling procedure; the position of the NW prior to any poling is marked by the dashed yellow line. This striking phenomenon was reproducibly observed only for down-poling procedures over the course of \tilda{}{10} poling experiments. Figures \ref{fig:movement}d,e show the NW topography \textit{during} poling, with the sudden movement of the NW during down-poling captured - as marked by the pink arrow. Figure \ref{fig:movement}f shows two PFM signal profiles taken along the NWs after the up- and down-poling procedures. Generally, the length of the poled regions is consistent with the poling procedures, however two observations are noteworthy: firstly, although the up-poling was done across \tilda{}4 $\mu$m, the measured PFM value is not uniform, with a strong response in the lower part of the poled region. {This probably resulted from a relatively fast scanning rate (7$\mu$m/sec) used in the up-poling scan. The dip observed towards the bottom end of the frame could result from double scanning by the poling tip, due to the start of a new frame.} Secondly, the area which was previously strongly poled in the up direction, became down-poled (black arrow in all sub-figures); we believe that this is a manifestation of a rotation movement of this part of the NW, resulting from the jumping movement of the down-poled region (pink arrow in all sub-figures), {effectively dragging the nearest parts of the NW along with it.} Other examples for NW movement are shown in Supporting Information (Fig.\nolinebreak\ S4), and in Fig.\nolinebreak\ \ref{fig:QNM} below. In order to verify that this is not simply a result of mechanical movement by the tip, a NW was imaged repeatedly, in order to induce mechanical effects. There was no corresponding macro-scale movement of the NW observed, only local damage (ripping) was imaged - see Supporting Information (Fig.\nolinebreak\ S5).\\ \indent
We now discuss the intrinsic mechanical properties of the poled NWs as measured by QNM. The experimental procedure was to pole sections of NWs as shown in Fig.\nolinebreak\ \ref{fig:poling}, and then image the same areas using QNM. Figure \ref{fig:QNM} shows a series of SPM images of the same NW, taken before and after up- and down-poling of separate segments of the NW; the upper sub-figures were taken during contact mode PFM scanning, while the bottom sub-figures were taken during QNM measurements, with a different AFM tip. The red (smaller) and green (larger) circles point out the same features on the substrate that are used as markers to ascertain the location of the features in the QNM measurement relative to the PFM measurement (arrows point to poled regions).  Notably, the feature in the green circle was observed only after the poling process, and was the residue left on the substrate as a result of the up-poling process 
(similar to what was shown in Fig.\nolinebreak\ \ref{fig:poling}). This region is located away from the NW, since the down-poling process resulted in a significant displacement of the NW relative to its original location (where the residue was found). Following the successful location of the poled regions (Fig.\nolinebreak\ \ref{fig:QNM}c), a slow, high-resolution scan (0.1 Hz, 1024 samples/line) was performed (Fig.\nolinebreak\ \ref{fig:QNM}d). A localized reduction of the extracted elastic modulus was observed corresponding to the poled regions of the NW, as is evident by the change of dominant colors from black-yellow (on the top and sides of the NW)  to green-orange. A line scan taken from that region (Fig.\nolinebreak\ \ref{fig:QNM}e) indicated a reduction of 500 MPa in the extracted modulus value. This observation is in agreement with previous reports of a lower modulus along the poled axis of P(VDF-TrFE)\cite{Roh2002}, although other reports indicate the poled orientation has a larger elastic modulus\cite{Otani2000}. It is important to note that while the Derjaguin-Muller-Toporov (DMT) model used by the QNM software to extract elastic moduli is suitable for planar samples but not directly to NWs, the qualitative comparison between the different regions of the same NW is still valid (see Supporting Information Sec.\nolinebreak\ IV for further discussion)\cite{QNM}.\\ \indent
A possible origin of the observed localized electromechanical interaction effects is in the inherent asymmetries in the chemical and electrical interactions of P(VDF-TrFE) in different polarization states with the substrate, enhanced by the additional mechanical degrees of freedom of the tip-NW-substrate system, as compared to the plate capacitor configuration, or the tip-TF-substrate system. Asymmetric effects related to polarization of PVDF based materials, and indeed ferroelectric materials in general, manifest as a difference in coercive voltage of up and down polarizations, or between subsequent poling steps (imprint effect), and are usually attributed to electrode work function and surface oxide, trapped interface charge, and film disorder \cite{Grossmann2002,Furukawa2006,Lew2009,Sharma2012}. Although critical to ferroelectric device performance, there is not yet a full understanding of these effects. Of these, we rule out the effect of electrode surface oxide, as this was shown to be considerably reduced when using gold electrodes (as in our case)\cite{Furukawa2006,Lew2009}. {However, Schottky effect might induce preferential (and self-) poling to the material due to the differences in work-function (see also SI}\cite{Kholkin1998,Choi2016}.\\ \indent
A basic distinction between the NW and the TF configurations is the non-clamped nature of the NW on the substrate, both laterally and relative to the substrate\cite{Yudin2001,Safari2008}. The absence of lateral PFM from the poled regions of the NW, compared to the clear, polarization dependent signal from the TF sample (Fig.\nolinebreak\ \ref{fig:poling} and Supp.\nolinebreak\  S3), is an interesting manifestation of this. This result is surprising, considering that P(VDF-TrFE) $\beta$-phase crystal structure results in the preference of 60$^{\circ}$ rotations of the horizontal dipole configuration (see Fig.\nolinebreak\ \ref{fig:scheme})\cite{Furukawa2006,Guo2013}. Therefore some lateral signal is expected during up- or down-poling, and its absence in the NWs is intriguing. In addition, the distinct nano-indentation of NWs and TFs also demonstrates this, as shown by finite-element simulations (see Supporting Information, Fig.\nolinebreak\ S6). If so, this lack of constraint is expected to allow relaxation of the material to more stable configurations, and indeed we observe changes in the extracted \textit{d}\sub{eff} values obtained from the NWs in the range of minutes, while these changes are notably reduced from the TF samples. {It might also account for the larger values usually measured (and in this work as well) for NW compared to thin films. When considering a NW section being poled, the energy needed to pole an equivalent clamped section would likely be higher than the energy needed to overcome the restriction imposed by the clamped configuration.}\\ \indent 
The lack of constraints may also explain the increased mechanical freedom of the NW system, wherein the asymmetries between the two polarization states are more pronounced. As mentioned above, the tip-NW-substrate is asymmetrical; in addition, the P(VDF-TrFE)/metal interface is inherently asymmetrical, as the molecular layer closest to the metal is different for each polarization, and for the pristine NW. Apart from work function related phenomena, one can consider the molecular nature of the interface. Figure \ref{fig:scheme} schematically shows the interface in different configurations: in case of up-poling, the atoms closest to the surface are a uniform layer of fluorine atoms (in the crystalline regions), while in the down-polarization the closest atoms are a blend of fluorine and hydrogen atoms; in the case of ``side" poling the molecular chains are stacked horizontally, with the dipole pointing horizontally, and the closest atoms to the metal are again fluorine and hydrogen. This implies that the chemical interactions as well as the electrical interaction of the interface molecules and dipoles are different for each configuration. The dipole-substrate interaction is thus dependent upon the dipole orientation and its distance from the surface, both of which have been shown to be very different for different polarizations\cite{Holmstrom1986,Cowin1994,Kokalj2011,Gabovich2012}. As an example, when considering dipoles perpendicular and parallel to a metallic surface, a single dipole would be more stable in a perpendicular orientation due to the attraction to its image, while an array of dipoles might be more stable parallel to the surface due to mutual repulsion of perpendicular dipoles\cite{Kokalj2011}. Furthermore, there could be a change in the preferred orientation of a single dipole (HF molecule in this case) from perpendicular to parallel, when approaching the surface\cite{Gabovich2012}.\\ \indent
If so, it is reasonable that the different polarization states have different energies on the gold substrate; in our case, for down-polarization, considering the higher \textit{d}\sub{eff} values obtained, and the shorter poling times (higher tip velocity which still enables poling), it appears that this polarization is considerably more stable compared to up-polarization. This observation is in agreement with the report by Sharma et al.\nolinebreak\ and Kim et al., where down-poled domains where found to grow more efficiently (with time as well as with voltage), even to an extent of exhibiting a non-linear increase in radius, after a certain threshold\cite{Sharma2012,Kim2010}. Interestingly, when considering single ions (having a much larger interaction with an adjacent metallic surface) DFT calculation show that H\super{+} is considerably more stable compared to F\super{-}\cite{Gabovich2012}, providing perhaps some insight into the energy consideration of equivalent layer termination in the case of P(VDF-TrFE). Dipole-substrate interaction was recently suggested as a driving force for the preferred edge-on, and increased crystallinity, of Pd-doped PVDF TFs\cite{Mandal2012}. A similar effect has also been reported for Ag-doped PVDF-TrFE\cite{Paik2015}.\\ \indent
The observation of the pristine sections of the NW exhibiting a consistent down-poled \textit{d}\sub{eff} value of about 5-6 pm/V, may further indicate that down-polarization is favourable in our system resulting from NW-substrate interaction as discussed above. Alternatively, the NW may possess a net, non-axial component of polarisation which, upon dispersion on the surface, aligns according to the interface preference. The AAO template related process should not result in an inherent asymmetry, therefore it is reasonable that the former case is prevalent; however, we cannot currently distinguish between the two alternatives, and future work will be concerned with this issue. The local damage observed in some of the up-poled regions may then be due to increased mechanical stresses induced by the presence of conflicting forces in these regions, resulting perhaps in amorphization or local loss of ferroelectric ordering, and thus the correspondingly reduced \textit{d}\sub{eff} extracted. This is also consistent with the observed reduced elastic modulus in this region, as the elastic modulus of polymers is expected to increase with crystallinity\cite{Holliday1975,Baniasadi2016}. The sudden $\mu$m-scale movement of the NW during poling (Fig.\nolinebreak\ \ref{fig:movement}) could be related to a non-linear response and the instantaneous expansion of the domain such that mechanical stresses resulting from the switch are released through this movement, which is otherwise not possible in TF samples.\\ \indent
Recently, Guo and Setter have developed a model\cite{Guo2013} based on bulk (or volume) dipole-dipole interactions to explain thickness dependent preferred orientations of P(VDF-TrFE) TFs. By summing up the interaction energies of the horizontal and the 60$^{\circ}$ rotated orientations (Fig. \nolinebreak\ \ref{fig:scheme}b), for a given volume, they point to a cross-over of the energies when the thickness of the TF is comparable with the grain size; thus inducing preference of the horizontal orientation for ``very" thin films\cite{Guo2013}. In contrast to TFs, NWs inherently have similar ``width" and ``thickness" (and in the case of \tilda{} 200 nm NWs, can accommodate symmetric grains\cite{Guo2013}), and so the volume model mentioned above would result in similar contributions of the different orientations to the bulk energy, thus effectively allowing for other effects, such as surface (or interface) effects, to come into play.\\ \indent
To conclude, we have performed localised poling experiments on single P(VDF-TrFE) NWs and on spin coated TFs of the same composition using an AFM tip. The results indicate significant distinctions in the electromechanical behavior of NWs and TFs; most notably was the absence of lateral PFM signal from poled sections of NWs, and a strong asymmetry between up- and down-poling efficiency in the NWs. On a larger scale, we have observed movement of the NW in exclusive relation to the down-poling procedure. We have subsequently examined, using QNM, the elastic properties of the poled and un-poled regions of the NWs, and found that the up-poled region became softer following poling.\\ \indent We attribute these observations to the different electronic and chemical surface interactions of the P(VDF-TrFE) with the substrate and/or ambient, which prevail due to the uniform geometry of the NW, and the reduced mechanical clamping which otherwise plays a significant role in TF samples. We believe that these result shed light on the electromechanical interplay in regards to ferroelectric P(VDF-TrFE) NWs, as well as on fundamental processes in this material that are not commonly observed when studying TFs.\\ \indent

\textbf{Supplementary Material}\\ \indent
See supplementary material for details of the scanning probe microscopy modes used in this work, materials synthesis, calibration procedure and data for PFM and QNM, additional control experimental data and computational modelling of local indentation in thin films and nanowires of P(VDF-TrFE).\\ \indent

\textbf{Acknowledgments}\\ \indent
S.K-N and Y.C are grateful for financial support from the European Research Council through an ERC Starting Grant (Grant no. ERC-2014-STG-639526, NANOGEN). R.A.W. thanks the EPSRC Cambridge NanoDTC, EP/G037221/1, for studentship funding. Q.J is grateful for financial support through a Marie Sklodowska Curie Fellowship, H2020-MSCA-IF-2015-702868.

\newpage
\end{document}